\documentclass[12pt]{article}
\usepackage{amsmath, amssymb, graphics}
\usepackage{setspace}
\usepackage[note]{achemso}
\usepackage{amsmath}
\usepackage{multicol}
\usepackage{subfigure}
\usepackage{graphicx}
\usepackage{rotating}
\usepackage{geometry}
\usepackage{verbatim}
\usepackage{multirow}
\usepackage{epstopdf}
\usepackage{comment}
\usepackage{amssymb}
\newcommand{\mathsym}[1]{{}}

\begin{document}

\title{Crowding Effects on the Mechanical Stability and Unfolding Pathways of Ubiquitin.}
\author{David L. Pincus$^{1}$ and D. Thirumalai$^{1,2,}$\footnote{Corresponding author phone: 301-405-4803; fax: 301-314-9404; thirum@umd.edu}}
\maketitle
\noindent $^1$Biophysics Program, Institute for Physical Science and Technology\\
$^2$Department of Chemistry and Biochemistry\\
$\text{  }$University of Maryland, College Park, MD 20742\\


\begin{abstract}
The interior of cells is crowded thus making it important to assess the effects of macromolecules on the folding of proteins. Using the Self-Organized Polymer (SOP) model, which is a coarse-grained representation of polypeptide chains, we probe the mechanical stability of Ubiquitin (Ub) monomers and trimers ((Ub)$_3$) in the presence of monodisperse spherical crowding agents.  Crowding increases the volume fraction ($\Phi_c$)-dependent average force ($\langle f_u(\Phi_c) \rangle$), relative to the value at $\Phi_c = 0$,  needed to unfold Ub and the polyprotein.   For a given $\Phi_c$, the values of $\langle f_u(\Phi_c) \rangle$ increase as the diameter ($\sigma_c$) of the crowding particles decreases.  The average unfolding force $\langle f_u(\Phi_c) \rangle$ depends on the ratio $\frac{D}{R_g}$,  where $D \approx \sigma_c (\frac{\pi}{6 \Phi_c})^{\frac{1}{3}}$ with $R_g$ being the radius of gyration of Ub (or (Ub)$_3$) in the unfolded state.  Examination of the unfolding pathways shows that, relative to $\Phi_c = 0$, crowding promotes reassociation of ruptured secondary structural elements.   Both the nature of the unfolding pathways and $\langle f_u(\Phi_c) \rangle$ for (Ub)$_3$ are altered in the presence of crowding particles with the effect being most dramatic for the subunit that unfolds last.  We predict, based on SOP simulations and theoretical arguments, that $\langle f_u(\Phi_c) \rangle \sim \Phi_c^{\frac{1}{3\nu}}$, where $\nu$ is the Flory exponent that describes the unfolded (random coil) state of the protein.
\end{abstract}

\section*{Introduction.}

Cells exist in a crowded environment consisting of macromolecules (lipids, mRNA, ribosome, sugars, etc.), making it critical to investigate protein folding in the presence of crowding agents \cite{MintonARB08}.  If the interactions between the crowding agents and the protein of interest are short-ranged and non-specific (as is often the case), then the volume excluded by the crowding agents prevents the polypeptide from sampling extended conformations.  As a consequence, the entropy of the denatured state ensemble (DSE) decreases relative to the case when the crowding agents are absent.  These arguments suggest that excluded volume of crowding agents should enhance the stability of the folded state provided that the crowding-induced changes in the native state are negligible \cite{Minton1:2005,Cheung:2005}.  The entropic stabilization mechanism, described above, has been used in several theoretical models to quantitatively describe the extent of folded protein as a function of the volume fraction, $\Phi _c$, of the crowding agents \cite{Cheung:2005,ZhouACR04}.  More recently, a theory whose origins can be traced to the concept of intra-protein attraction due to depletion of crowding agents near the protein \cite{Asakura:1954, Asakura:1958, Vrij:1976}, predicts that the enhancement in stability, $\Delta T(\Phi _c) = T_f(\Phi _c) - T_f(\Phi _c = 0) \sim \Phi _c^{\alpha}$, where $T_f(\Phi _c)$ is the folding temperature at $\Phi _c$ and $\alpha$ is related to the Flory exponent that characterizes the size of the protein in the DSE \cite{Cheung:2005}.  From this prediction it follows that crowding affects the DSE to a greater extent than the folded state.  Although the precise theoretical predictions of the power law change in $\Delta T(\Phi _c)$ as $\Phi _c$ changes have not been verified, several experiments using a number of proteins have confirmed that indeed $T _f(\Phi _c)$ increases with $\Phi _c$ \cite{MintonJMB03,CheungPNAS07,CheungPNAS08}.  It cannot be emphasized enough that the theory described here applies only to cases when the crowding interactions between crowding agents and proteins and between crowding particles themselves are purely repulsive.

While much less is known about the effects of crowding on the folding kinetics, Cheung et al.\ \cite{Cheung:2005} predicted that the entropic stabilization also suggests that the folding rates should increase at moderate values of $\Phi _c$.  They suggest that crowding can enhance folding rates by a factor $e^{\Delta S(\Phi _c)/k_B}$ where $\Delta S(\Phi _c)$ $(\sim \Phi_c^\alpha)$ is the decrease in the entropy of the DSE relative to its value in the bulk.  From the arguments of Cheung et al.\ \cite{Cheung:2005} it follows that the equilibrium changes in the entropy ($\Delta S(\Phi _c)$) of the DSE, with respect to the bulk, should also determine rate enhancement provided that neither the barriers to folding \cite{CheungJPCB07} nor the native state is perturbed significantly by crowding particles.

Single molecule force spectroscopy, such as Atomic Force Microscopy (AFM) and Laser Optical Tweezers, have been used to monitor the behavior of biopolymers under tension are ideally suited to probe the enhancement in crowding-induced stability by a direct measurement of $f_u(\Phi _c)$.  Indeed, Ping et al.\ \cite{Ping:2006} have recently investigated the effect of Dextran molecules on the mechanical stability of $(\text{Ub})_8$.  The 8 Ub (Fig.\ \ref{fig:1}{\bf A}) modules were N-C linked (i.e., modules i and i+1 were chemically linked together in a head-to-tail manner).  
They found that the average force required to unfold a module, $\langle f_u(\Phi _c) \rangle$, increased by 21$\%$ as the Dextran
concentration, $\rho $, was increased from 0 to 300 g/L at $r_f = 4.2 \times 10^3$ pN/s. 
Similar results have been obtained recently by Yuan et al.\  at $r_f = 12.5 \times 10^3$ pN/s \cite{Yuan:2008}.

Motivated in part by experiments \cite{Ping:2006,Yuan:2008}, we used simulations to investigate the effects of crowding agents on the mechanical
stability of a protein subject to external tension.  We focused on ubiquitin (Ub), a 76-residue protein composed of 5 $\beta
$-strands and 2 $\alpha $-helices (Fig.\ \ref{fig:1}{\bf A}), and confined our investigation to non-equilibrium `force-ramp'
experiments \cite{Fisher:2000}.  The primary data recorded during such an experiment is a trace of the force exerted on the tip as a function of the
extension of the molecule; a force-extension curve (FEC).  When the force exceeds some critical value, the FEC displays a sudden increase
in length and is often accompanied by a concomitant sharp decrease in force.  Presumably, the sharp change corresponds to the unfolding of the
protein.  Typical AFM experiments use tandem arrays of proteins which are chemically linked together (often through genetic engineering).  We use the term module to denote a protein of the array.  The FEC resulting from such an experiment reveals several equally spaced peaks punctuated by sharp increases in the extension of the molecule corresponding to the unfolding
of individual modules.  The height of these force peaks and their shape depend on the loading rate, $r_f = k_s \times \mathit{v}$,
where $k_s$ is the cantilever's spring constant and $\mathit{v}$ is the (constant) speed at which the stage is retracted away from the
cantilever \cite{Evans:1997}.

In order to compare to experiments our simulations are performed using coarse-grained models for which simulations can be done at $r_f$ that are comparable to those
in AFM experiments.  Our work has led to a number of testable results: (1) At $\Phi _c = 0.3$ the average unfolding force for Ub increases by at most only 7\% compared to $\Phi _c = 0$.  We find that $\langle f_u(\Phi _c) \rangle$ in small crowding agents is greater than in larger particles.  (2)  In the presence of crowding agents, secondary structural elements reform multiple times even after initial rupture.  (3)  Although large crowding particles are predicted to have a smaller effect on $\langle f_u(\Phi _c) \rangle$ (for a given $\Phi _c$), they can profoundly affect the unfolding of poly Ub.  We predict that $\langle f_u(\Phi _c) \rangle$ for a given subunit depends on the number of already unfolded portions of the poly protein.  This result is important because many naturally occurring proteins that are subject to tensile stresses exist as tandem arrays of modules.  It further suggests that the existence of such redundancy can more properly be understood in the context of a crowded cellular milieu.

\section*{Methods.}

\subsection*{ Self-Organized Polymer Model for Ub.}

We used a coarse-grained model for proteins to investigate crowding effects on the mechanical stability of Ub and $(\text{Ub})_3$ at loading rates that are comparable to those used in AFM experiments \cite{Ping:2006,Yuan:2008}.  We assumed that Ub could be described using the Self-Organized Polymer (SOP) model; a model that has been successfully used to make a number of predictions regarding the unfolding of proteins and RNA \cite{Hyeon:2006,Hyeon2:2006}, allosteric transitions in enzymes \cite{Hyeon3:2006,Chen:2007}, and movement of molecular motors on polar tracks \cite{Hyeon2:2007}.  Previous studies \cite{BestJPCB08} have used more standard Go-models \cite{KaranicolasProtSci02,Li:2007} to probe various aspects of forced unfolding of Ub.  The SOP energy function ($E_p$) for a protein with N amino-acids, specified in terms of the C$_\alpha $ coordinates $\mathbf{r}_i$ $(i = 1, 2,..., N)$, is
\begin{equation}\label{eqn:1}
\begin{split}
E_p 
&= E_{FENE} + E_{nb}^{att} + E_{nb}^{rep} \\
&= - \sum _{i=1}^{N-1} \frac{k}{2} R_0^2\ln \left[1-\frac{\left(r_{i,i+1}-r_{i,i+1}^0\right)^2}{R_0^2}\right]+\sum _{i=1}^{N-3}
\sum _{j=i+3}^N \varepsilon _h \left[\left(\frac{r_{ij}^0}{r_{ij}}\right)^{12}-2\left(\frac{r_{ij}^0}{r_{ij}}\right)^6\right]\Delta
_{ij}\\
&+\sum _{i=1}^{N-2} \sum _{j=i+2}^N \varepsilon _l\left(\frac{\sigma }{r_{ij}}\right)^6\left(1-\Delta _{ij}\right),
\end{split}
\end{equation}
\noindent where $r_{ij} = | \mathbf{r}_i-\mathbf{r}_j |$, $r_{ij}^0 = | \mathbf{r}_i^0-\mathbf{r}_j^0 |$ is the value of
$r_{ij}$ in the native structure, $k = 2 \times 10^3$ kcal/(mol$\cdot $\(\text{nm}^2\)), $\varepsilon _h = 1.4$ kcal/mol, $\varepsilon _l = 1.0$ kcal/mol, and $\sigma = 0.38$ nm.  
Note that $k_BT \approx 0.6$ kcal/mol $\approx 4.2$ pN$\cdot$nm.  In Eq.\ (\ref{eqn:1}) $\Delta _{ij} = 1$ if $r_{ij}^0 < 0.8$ nm, and $\Delta _{ij} = 0$ otherwise.  
Native coordinates corresponded to 
those of the C$_\alpha $ atoms of the 1.8 {\AA} resolution Protein Data Bank crystal structure 1UBQ \cite{Vijay-Kumar:1987}.  For Ub $N = 76$ and $N
= 228$ for $(\text{Ub})_3$.  The first term in Eq. (\ref{eqn:1}) is the FENE potential \cite{Kremer:1990} that accounted for chain connectivity.  
The second (Lennard-Jones) term accounted for the non-bonded interactions that stabilize the native state, and the final (soft-sphere) term accounted for excluded-volume interactions
(including those of an angular nature).  The SOP model is different from the Go-model because there are no angular terms in SOP, and the connectivity is enforced differently as well.  The SOP 
representation of the polypeptide chain is in the same spirit as other coarse-grained models used in polymers \cite{pincusMinMod08}.

\subsection*{ Crowding Particles and Interactions with Ub.}

We assumed that the crowding particles are spherical with diameter $\sigma _c$. ($\sigma _c = 6.4$ nm in some simulations, while $\sigma _c = 1.0$ nm in others.)  
Crowders interacted amongst themselves and with the protein, respectively, via the following LJ potentials:
\begin{equation}\label{eqn:2}E_{\text{cc}}= 4\varepsilon_l \left(\left(\frac{\sigma _{\text{cc}}}{r}\right)^{12}-\left(\frac{\sigma _{\text{cc}}}{r}\right)^6+\frac{1}{4}\right)\Theta
\left(r_{\text{min}}^{\text{cc}}-r\right)\end{equation}

\begin{equation}\label{eqn:3}E_{\text{cp}}= 4\varepsilon_l \left(\left(\frac{\sigma _{\text{cp}}}{r}\right)^{12}-\left(\frac{\sigma _{\text{cp}}}{r}\right)^6+\frac{1}{4}\right)\Theta
\left(r_{\text{min}}^{\text{cp}}-r\right),\end{equation}
\noindent where $\sigma _{\alpha \beta } = (\sigma _{\alpha } + \sigma _{\beta })/2$, $\sigma _p = \sigma = 0.38$ nm, $r_{\text{min}
}^{\text{cc}} = 2^{1/6}\sigma _{\text{cc}}$, and $r_{\text{min} }^{\text{cp}} = 2^{1/6}\sigma _{\text{cp}}$.  The Heaviside functions
truncate the potentials at their minima and thereby ensured that only the repulsive portions of Eqs.\ (\ref{eqn:2}) and (\ref{eqn:3}) contributed to interactions involving crowding agent. 

\emph{Mimics of Crowding Using Asakura-Oosawa Theory:}
Even using a coarse-grained SOP representation of proteins, it is difficult to carry out converged simulations in the presence of crowding agents.  The reason is that the number of crowding agents can be large.  Moreover, the separation in the spatial and temporal scales of the protein and the crowding particles has to be carefully considered to obtain reliable results.  In light of these difficulties, it is of interest to consider the effective attraction between the sites on the protein using the implicit pairwise potential computed by Asakura and Oosawa.  The intramolecular attraction arises due to the depletion of crowding particles near the protein.  To probe the efficacy of these models we employed in some simulations the Asakura-Oosawa model \cite{Asakura:1954, Asakura:1958, Vrij:1976} of crowding effects.  For these simulations, we added the following term to the bare SOP Hamiltonian (Eqn. (\ref{eqn:1})):
\begin{equation}\label{eqn:8}E_{\text{AO}}\left(r_{ij}\right)=-\Phi _c  k_BT \sum _{j\geq i+3} \left(\frac{\left(\sigma +\sigma _c\right)}{\sigma _c}\right)^3\left(1-\frac{3
r_{ij}}{2 \left(\sigma +\sigma _c\right)}+ \frac{r_{ij}^3}{2 \left(\sigma +\sigma _c\right)^3}\right)\text{                                               }\sigma < r < \sigma +\sigma _c,\end{equation}
\noindent where $\sigma = 0.38$ nm, $\sigma _c = 6.4$ nm, $\Phi _c = 0.3$, $k_BT \simeq 4.2$ pN$\cdot$nm, and $r_{ij}$ is the distance separating protein beads $i$ and $j$.

\subsection*{ $(\text{Ub})_3 $ Intermodule Interactions.}

For simulations involving $(\text{Ub})_3$, residues in different modules interacted via:

\begin{equation}\label{eqn:4}E_{\text{pp}} = \varepsilon_l  \left(\frac{\sigma }{r}\right)^6,\end{equation}

\noindent where r is the distance separating the two beads.  Note that this potential is
short-ranged and purely repulsive and that it is the same potential used for non-native intra-protein interactions \cite{Hyeon2:2006}.

\subsection*{ Simulation Details.}

\emph{$\Phi _c = 0$}: Hundreds of simulations of $5 \times 10^6$ steps ($\simeq 30 \mu$s) at $T = 300$ K were used to generate initial structures for use in the
pulling simulations.  The protein was completely free in solution (i.e., no forces were applied to either terminus), and no crowders were present
during the equilibrations.  The N-terminus of the protein was subsequently translated to the origin, and the protein was rotated such that its
end-to-end vector, $\mathbf{R}$, (i.e., the vector pointing from the N-terminal bead to the C-terminal bead) coincided with
the pulling (+z) direction.

An unfolding trajectory was initiated by selecting a random initial structure from amongst the set of thermally equilibrated structures, and tethering a harmonic spring to the C-terminal bead.  
The N-terminal bead was held fixed throughout the simulations.  Tension was applied to the protein by displacing the spring along the +z
axis and resulted in application of the following force to the C-terminal bead:

\begin{equation}\label{eqn:5}f_z = -k_s\left([z(t)-z(0)]-\left[z_s(t)-z_s(0)\right]\right), \end{equation}

\noindent where $k_s$ is the spring constant, $z(t) = \mathbf{R}(t)\cdot\overset{\wedge }{\mathbf{z}}$, and $z_s(t)$ corresponds
to the displacement of the end of the spring.  Note that $k_s$ was also used to constrain the simulation to the z-axis; $f_x = -k_s[x(t)-x(0)]$ and 
$f_y = -k_s[y(t)-y(0)]$.  The displacement of the spring was updated at every timestep.

We simulated forced-unfolding of monomeric Ub at four different $r_f$ ($160 \times 10^3$ pN/s, $80 \times 10^3$ pN/s, $20 \times 10^3$ pN/s, 
and $4 \times 10^3$ pN/s), while simulations on N-C-linked $(\text{Ub})_3$ were performed at $r_f = 640 \times 10^4$ pN/s.  
All overdamped force-ramp simulations were performed at the same speed $\mathit{v} = 10312$ nm/s, 
and spring constants were varied over a range from 0.3879 pN/nm - 31.032 pN/nm to achieve the aforementioned $r_f$ (via the relation $r_f = k_s\mathit{v}$).  
Our simulations were realistic because they maintained loading rates consistent with experiment and because $r_f$ is the prime determinant of unfolding pathway \cite{Evans:1997}.

\noindent \emph{$\Phi _c \neq$ 0.0}: Simulations involving explicit crowders were carried out at a fixed volume fraction $\Phi _c = 0.3$ and with a fixed number $N_c = 100$ of crowding
spheres.  ($N_c$ was fixed to render the problem computationally tractable).  Using the relation $\Phi _c = \frac{N_c\pi}{6}\left(\frac{\sigma_c}{L}\right)^3$, 
we adjusted the length of a side of the cubic simulation box ( $L$ ) to maintain $\Phi _c = 0.3$.  Thus, $L = 35.8$ nm when $\sigma
_c = 6.4$ nm and $L = 5.6$ nm when $\sigma _c = 1.0$ nm.  Explicit crowders were added to the simulation after loading an equilibrated structure but before the application of tension.  
Initial crowder positions were chosen randomly and in a serial manner from a uniform distribution.  If the distance between an initial crowder position
and that of another crowder or protein bead did not exceed the sum of their radii, then the prospective position was rejected and another random position chosen 
to avoid highly unfavorable steric overlaps.

Periodic boundary conditions (PBC) and the minimum image convention \cite{Allen:1987} were employed in the simulations.  Two
sets of coordinates were stored for protein beads at every timestep; PBC were applied to one set and the other was propagated without PBC.  Distances
between protein beads were calculated from the uncorrected set of coordinates without minimum imaging, while protein-crowder distances were calculated
from the PBC coordinates with minimum imaging.

To improve simulation efficiency, a cell list \cite{Allen:1987} was used to calculate crowder-crowder and protein-crowder interactions.  The
entire simulation volume was partitioned into 64 subvolumes, and it was only necessary to calculate interactions within a subvolume and between beads
of the subvolume and those of 13 of its 26 neighbors.  The cell-list was updated at every timestep to ensure the accuracy of the simulations.

The equations of motion in our overdamped simulations (used in all force-ramp simulations) were integrated with a timestep $h = 0.01 \tau _L$ ($\tau _L = 2.78$ ps) 
using the method of Ermak and McCammon \cite{Ermak:1978}.  The friction coefficient of the crowders, $\zeta _c$, was determined
via the relation $\frac{\zeta _c}{\zeta } = \frac{\sigma _c}{\sigma }$, where $\sigma _c$ is the crowder diameter, $\sigma = 0.38$ nm is the
diameter of a protein bead, and $\zeta = 83.3 \times (\frac{m}{\tau _L}) = 9 \times 10^{-9}$ g/s is the friction coefficient associated
with a protein bead of mass $m = 3 \times 10^{-22}$ g.  Simulated-times were translated into real-times using 
$\tau _H = (\frac{\zeta \epsilon _h}{k_BT}) \times \tau _L \times (\frac{\tau _L}{m})$ \cite{Veitshans:1996}.  At $T = 300$ K, $\tau _H = 543.06$ ps, and since $h = 0.01 \times \tau _L$ 
the real-time per step is 5.4306 ps.

\section*{Results and Discussion.}

\subsection*{ Monomeric Ub at $\Phi _c = 0.0$.}

At $\Phi _c = 0.0$, forced unfolding of Ub was simulated at four different $r_f$ ($4 \times 10^3$ pN/s, 
$20 \times 10^3$ pN/s, $80 \times 10^3$ pN/s, $160 \times 10^3$ pN/s), where the lowest value corresponds approximately
to the value used in the pulling experiments of Ping et al.\ \cite{Ping:2006}, and all $r_f$ are experimentally accessible.

\emph{Force Profiles}: Fig.\ \ref{fig:2}({\bf A} and {\bf B}) provides examples of FEC's collected at the highest and lowest $r_f$.
We used a nominal contour length of ($N-1$) $\sigma = 75 \times 0.38$ nm and unfolding forces, $f_u$, to determine contour-length increments, $\Delta \mathcal{L}$, for each trajectory
at $r_f = 160 \times 10^3$ pN/s (Fig.\ \ref{fig:2}{\bf A}).  We identified $f_u$ with the peak
of the FEC before the stick-slip transition \cite{Carrion-Vazquez:2003,Bockelmann:1998}. The average extension $\langle \Delta
\mathcal{L} \rangle = 23.991 \pm 0.010$ nm is in excellent agreement with the experimental result of 24 $\pm $ 5 nm found by
Carrion-Vazquez et al.\ \cite{Carrion-Vazquez:2003}.  The projection ($z_u$) of the end-to-end vector at $f_u$ in the z-direction varied between 4.1 nm and 4.7 nm, depending on $r_f$.  Since the native end-to-end distance, $z_0 = 3.7$ nm, $z_u - z_0 \equiv \Delta z_u$ ranges from 0.4-1.0 nm.  The lower end of this range is slightly larger than the 0.25 nm transition-state
distance for the mechanical unfolding of the structurally similar titin immunoglobulin domains \cite{Carrion-Vazquez:1999}.  Indeed,
we expect $\Delta z_u > 0.25$ nm, because of the non-equilibrium nature of the simulations.  Larger $r_f$ typically lead to larger $\langle \Delta z_u \rangle$ ($\sim k_BT / \langle f_u \rangle \ln(r_f)$).

Average unfolding forces, $\langle f_u(\Phi _c) \rangle $, depended approximately logarithmically on $r_f$\cite{Evans:1997} (Fig.\ \ref{fig:2}{\bf C}), and $\langle f_u(r_f = 4 \times 10^3 \text{pN/s}) \rangle = 136$ pN is in fair agreement with the
experimental value of $166 \pm 33$ pN observed by Ping et al.\ \cite{Ping:2006} at $r_f = 4.2 \times 10^3$ pN/s.  The $\langle z_u \rangle$ 
also showed a logarithmic dependence on $r_f$, 
but the difference between the value calculated at $r_f = 160 \times 10^3$ pN/s and that calculated at $r_f = 4 \times 10^3$ pN/s is small ($\simeq $ 2 {\AA}).  Although
the underlying free-energy landscape is time-dependent in a non-equilibrium force-ramp pulling experiment, these results suggest that the distance
from the native state to the transition state is small.  It is likely that at loading rates that are achieved in laser optical tweezer experiments ($\sim$ 10 pN/s), the location to the transition state would increase because the response of biopolymers to loading rate changes from being plastic (low $r_f$) to brittle (high $r_f$) \cite{Hyeon:2007}. 

\emph{Unfolding Pathways}: In the dominant pathway, unfolding proceeded in a fairly Markovian fashion with the primary order of events following the sequence $\beta $1/$\beta $5 $\rightarrow $ $\beta $3/$\beta $5 $\rightarrow $ $\beta $3/$\beta $4 $\rightarrow $
$\beta $1/$\beta $2 (Figs.\ \ref{fig:1}{\bf B} and \ref{fig:12}). This is precisely the same sequence seen in the simulations by Li et al.\cite{Li:2007}.  
Alternative pathways, reminiscent of kinetic partitioning \cite{GuoBP95} observed in forced-unfolding of GFP \cite{Mickler:2007} and lysozyme \cite{Peng:2008}, were also infrequently sampled.  
For example, at $r_f = 4 \times 10^3$ pN/s, $\sim 6\%$ of the trajectories unfolded
as follows: $\beta $1/$\beta $5 $\rightarrow $ $\beta $1/$\beta $2 $\rightarrow $ $\beta $3/$\beta $5 $\rightarrow $ $\beta $3/$\beta $4 (Fig.\ \ref{fig:1}{\bf B}), while the remaining
$\sim 94\%$ followed the dominant pathway.  At the highest loading rate, one frequently observed the following sequence of events $\beta $1/$\beta
$5 $\rightarrow $ $\beta $3/$\beta $5 $\rightarrow $ $\beta $1/$\beta $2 $\rightarrow $ $\beta $1/$\beta $2 (reform) $\rightarrow $ $\beta $3/$\beta
$4 $\rightarrow $ $\beta $1/$\beta $2, where $\beta $1/$\beta $2 ruptured but then reformed prior to the rupture of $\beta $3/$\beta $4.  As illustrated in Fig.\ \ref{fig:2}({\bf A} and {\bf B}), unfolding events at smaller $r_f$ tended to result in larger molecular extensions.  

The non-equilibrium character of a pulling experiment decreased with decreasing $r_f$, and smaller $r_f$ resulted in smaller force-drops  after the unfolding force, $f_u$, is reached.  
Since the force applied by the spring to the end of the protein did not fall off as sharply at lower $r_f$, more of the protein was extended during an unfolding event.  The smaller extensions following unfolding events at higher $r_f$ (relative to those observed
at lower $r_f$) were responsible for the $\beta $1/$\beta $2 unfolding/refolding events mentioned above because the applied tension was very low after the initial rupture event (Fig.\ \ref{fig:2}({\bf A} and {\bf B})).  At lower $r_f$, this situation no longer held because the initial unfolding event resulted in a chain extension that was a significant fraction of the chain's contour length (Fig.\ \ref{fig:2} {\bf B}).

\subsection*{ Crowding Effects on Ub ($\Phi _c = 0.3$).}

Depletion forces stabilize proteins and shift the folding equilibrium towards more compact states \cite{Shaw:1991}.  These forces result from an increase in
the entropy of the crowding agents that more than compensates for an increase in the free-energy of a protein molecule upon compaction.  
Simulations of forced-unfolding of Ub in the presence of explicit crowders of diameters $\sigma _c = 6.4$ nm and $\sigma _c = 1.0$ nm were used 
to assess the contribution of the depletion forces to mechanical stability.  Sixteen trajectories were collected for each $r_f$ investigated. Three $r_f$ ($20 \times 10^3$ pN/s,
$80 \times 10^3$ pN/s, $160 \times 10^3$ pN/s), were explored for the $\sigma _c = 6.4$ nm sized depletants.  Only the two highest 
$r_f$ ($80 \times 10^3$ pN/s, $160 \times 10^3$ pN/s), were explored for the $\sigma _c = 1.0$ nm sized crowders.

\emph{Small crowding particles increase unfolding forces}:  Example FEC's collected in the presence of crowders
of diameter $\sigma _c = 6.4$ nm and $\sigma _c = 1.0$ nm are illustrated in Fig.\ \ref{fig:9}({\bf A} and {\bf B}).  
The $\sigma _c = 6.4$ nm curves in Fig.\ \ref{fig:9} look qualitatively very similar to those seen at $\Phi _c = 0.0$, while the
FEC's collected at $\sigma _c = 1.0$ nm look qualitatively different.  For example, larger $\langle f_u \rangle$ than those observed
at either $\Phi _c = 0.0$ or in the presence of the $\sigma _c = 6.4$ nm crowders are apparent in the FEC's collected at $\sigma _c = 1.0$ nm (Fig.\ \ref{fig:9}({\bf A} and {\bf B})).  
Indeed, Fig.\ \ref{fig:9}{\bf C} reveals that this observation is quantitatively accurate.  Although the $\langle f_u \rangle$ 
in the presence of the $\sigma _c = 6.4$ nm crowders were statistically indistinguishable from the $\langle f_u \rangle$ at $\Phi _c = 0$ 
(compare Fig.\ \ref{fig:9}{\bf C} with Fig.\ \ref{fig:2}{\bf C}), the average unfolding forces in the presence of the $\sigma
_c = 1.0$ nm crowders were statistically greater than those measured at $\Phi _c = 0.0$.  At $r_f = 160 \times 10^3$ pN/s, the $\langle f_u(\Phi _c = 0.3) \rangle$ 
in the presence of the $\sigma _c = 1.0$ nm crowders exceeded that at $\Phi _c = 0.0$ by 3$\%$, while at $r_f = 80 \times 10^3$ pN/s 
the increase was 4$\%$ (compare Fig.\ \ref{fig:9}{\bf C} with Fig.\ \ref{fig:2}{\bf C}).  
In one respect these results are not surprising; it follows from the AO theory and Eq. (\ref{eqn:6}) that smaller crowders stabilize Ub more than larger ones.  On the other hand,
the extent of stabilization (as measured by increases in $\langle f_u \rangle$) was small.  We should emphasize that although the increase in the unfolding force is small, the 
stability change upon crowding is significant.  The enhancement in stability is $\Delta G \sim \langle f_u(\Phi _c) \rangle\langle z_{DSE} \rangle \sim 5$ $k_BT$ using a 3\% increase in the unfolding force, and $\langle z_{DSE} \rangle$ the location of the unfolded basing $\approx 5$ nm.

\emph{Crowding leads to transient local refolding}:  The unfolding pathways in the presence of crowding agents of both sizes were very similar to those seen at $\Phi _c = 0.0$.  
Nevertheless, at $\Phi _c = 0.3$ there tended to be more unfolding/refolding
events as the molecule extended past $z_u$ (Fig.\ \ref{fig:12}) than at $\Phi _c = 0.0$. As illustrated, 
strand-pairing between $\beta $1 and $\beta $2 and between $\beta $3 and $\beta $4 persisted to a greater extent after the initial unfolding event at
$\Phi _c = 0.3$ than at $\Phi _c = 0.0$.  
As Ub passed through the point ($z_u$,$f_u$), its termini were extended to distances greater than the diameter of even the larger crowders.  Depletion forces resulting from the presence
of 6.4 nm crowders act on these larger length scales.  Assuming that the tension applied to the C-terminus is small enough
(e.g., after rupture events at higher loading rates), then such depletion forces can promote reformation of contacts between secondary structural
elements several times during the course of a trajectory.  If this is indeed the case, then it suggests that unfolding polyUb may be 
different from the unfolding of monomeric Ub, because depletion effects should increase with number of modules in the tandem (see below). 

\emph{AO Model For Forced-Unfolding in the Presence of Crowders}:  We used the Asakura-Oosawa AO model \cite{Asakura:1954, Asakura:1958, Vrij:1976} (Eq.\ (\ref{eqn:8})) of the depletion interaction to model the effects of a crowded environment on Ub.  The AO theory has been successfully used to model the effects of a crowded environment on polymers and colloids \cite{Asakura:1958,Vrij:1976,Verma:1998,Toan:2006}, and we found that it gives qualitatively accurate results for Ub.  We used Eq.\ (\ref{eqn:8}) to approximate the effective interaction between spherical protein beads immersed in a crowded solution of volume fraction $\Phi _c = 0.3$.  From the form of the AO potential between
protein beads, it follows that the range of the potential is proportional to $\sigma _c$, but the strength is greater for smaller crowders.  Indeed, as revealed in the previous section, simulations in the presence of explicit crowders showed that 1.0 nm crowders 
resulted in larger average unfolding forces than 6.0 nm crowders (Fig.\ \ref{fig:9}{\bf C}).

Although the AO-potential yields qualitatively accurate results, use of the AO-potential of Eq.\ (\ref{eqn:8}) to implicitly model non-bonded interactions in Ub did not yield quantitatively accurate results.  At $r_f = 160 \times 10^3$ pN/s $\langle f_u(\Phi _c = 0.0) \rangle = 175.78 \pm
1.52$ pN, while $\langle f_u(\Phi _c = 0.3) \rangle = 266.26 \pm  2.70$ pN.  Thus, simulations
with the AO potential led to a mean unfolding force that is roughly 50$\%$ greater than in its absence.  This disagrees sharply with the $\langle f_u(\Phi _c = 0.3) \rangle = 173.64 \pm 3.49$ 
pN resulting from our own simulations in the presence of explicit crowding agent at 
$\Phi _c = 0.3$ at $r_f = 160 \times 10^3$ pN/s ( Fig.\ \ref{fig:9}{\bf C} ).  Indeed, the result also stands in marked contrast
to the experimental results of Ping et al.\ \cite{Ping:2006} on octameric Ub, which saw a maximum increase in $\langle f_u \rangle$ of 21$\%$ (at $\Phi _c > 0.3$, $r_f = 4200$ pN/s, 
and with $\sigma _c \simeq 7.0$ nm) over the $\langle f_u \rangle = 166$ pN at $\Phi _c = 0$ \cite{Ping:2006}.

There are a couple of origins to the discrepancy.   First, the AO-potential was derived to understand the
equilibrium of colloidal spheres and plates in the presence of smaller-sized spherical crowding agents.
Our experiments were of a non-equilibrium nature, so it is somewhat unreasonable to expect
such simulations to yield quantitatively accurate unfolding forces or dynamics.  Second, as 
pointed out by Shaw and Thirumalai  \cite{Shaw:1991}, three-body terms are required to properly
model depletion effects even in good solvents let alone in the poor-solvent conditions of our
simulations.  To elaborate, let us consider the volume excluded to crowders by a Ub molecule, V$_{ex}$(Ub), to be the volume enclosed by a union of spheres 
of radii $S_i$ = $\frac{ \sigma+\sigma _c}{2}$.  With an AO potential, the volume excluded to the spherical crowding agents is: 

\begin{equation}\label{eqn:9}V_{E}\text{(Ub)}\simeq \sum _{i=1}^{ N} V(S_i)-\sum _{j>i} V(S_i\cap S_j),\end{equation}

\noindent where $N$ is the number of residues in monomeric Ub (76), $V(S_i)$ is the volume of $S_i$ and $V(S_i\cap S_j)$ is the volume associated with the
overlap of $S_i$ and $S_j$.  Eq.\ (\ref{eqn:9}) neglects the overlap of three or more spheres.  The importance of such overlaps, for soft-spheres, increases as the crowders become much larger than the protein beads, and the neglect of such overlaps is the reason for the quantitative inaccuracy.  As the size of the
crowders increases (i.e., as the thickness of the depletion layer surrounding the protein beads increases), the surface enclosing Ub becomes more spherical, loses detail, and undoubtedly changes less in response to changes in the conformation of the molecule.  Since depletion forces are proportional to the change in $V_{E}$(Ub) with respect to changes in Ub conformation, Eq.\ (\ref{eqn:9}) overestimates the size of depletion forces in the presence of large crowders.  Despite these limitations the AO model, which is simple, can be used to provide qualitative predictions.  Finally, we note that experiments typically use polyproteins to study force induced unfolding (e.g., Ping et al.\ \cite{Ping:2006} used (Ub)$_8$ in their experiments).  For polyproteins the 
quantitative accuracy of the AO theory for unfolding in the presence of crowders of diameter $\sigma_c = 6.4$ nm is likely to increase, because the volume excluded to 
the crowders will change significantly with changes in the conformation of the polyprotein.

\subsection*{ $(\text{Ub})_3$ at $\Phi _c = 0.0$ and $\Phi _c = 0.3$.}

From arguments based on volume exclusion that lead to crowding-induced entropic stabilization of the folded structures, it follows that crowding effects should be more dramatic on poly Ub than the monomer.  In order to illustrate the effect of crowding on stretching of (Ub)$_3$ we chose $\sigma _c = 6.4$ nm, which had negligible effect on 
$\langle f_u(\Phi_c) \rangle$ for the monomer.  However, we found significant influence of the large crowding particles when (Ub)$_3$ was forced to unfold in their presence.  
The FEC (Fig.\ \ref{fig:14}) shows three peaks that corresponded to unfolding of the three domains, when simulations were performed
at crowder volume fractions $\Phi _c = 0.0$ and $\Phi _c = 0.3$ at $r_f = 640 \times 10^4$ pN/s.  
Fig.\ \ref{fig:13} presents average unfolding forces as a function of the unfolding
event number.  Although the first two unfolding events were statistically indistinguishable at $\Phi _c = 0.0$ and $\Phi _c = 0.3$, the final event
occurred at much larger $\langle f_u \rangle$ in the presence of crowders than in their absence. 

{\it Order of Unfolding was Stochastic:}  $(\text{Ub})_3$ has 3 chemically identical modules.  In the pulling simulations the N-terminus 
of module $A$ was held fixed, while force was applied to the C-terminus of module $C$.  Figure \ref{fig:20} illustrates the time-dependence of contacts
between secondary structure elements of modules $A$, $B$, $C$.  It is clear from Fig.\ \ref{fig:20}{\bf A} that module $C$ unfolded
first, followed by module $B$, and finally by module $A$.  On the other hand, the order of events in Fig.\ \ref{fig:20}{\bf B} was $A \rightarrow C \rightarrow B$.  
The frequency with which the $3! = 6$ possible permutations of these orders at $\Phi _c = 0$ and at $\Phi _c = 0.3$ were observed is presented in 
Table \ref{tbl:moduleOrderFrequencies}.  It is clear that at both $\Phi _c = 0$ and at $\Phi _c = 0.3$, the most probably order of events was $C \rightarrow B \rightarrow A$.  
At $\Phi _c = 0$ this order was only marginally more probable than the order $C \rightarrow A \rightarrow B$, while at $\Phi _c = 0.3$ $C \rightarrow B \rightarrow A$ 
became overwhelmingly more probable than any other unfolding order.  In none of the simulations at $\Phi _c = 0$ or $\Phi _c = 0.3$, did the rupture of module $C$ occur as 
the final event.

{\it Unfolding Within a Module Depended on Proximity to the Point of Force Application:}  At $r_f = 640 \times 10^4 pN/s$, $(\text{Ub})_3$ is fairly brittle, and
the rupture of contacts within a module occured nearly simultaneously.  Nevertheless, by carefully examining time-dependent contact maps such as
those illustrated in Fig.\ \ref{fig:20}, we were able to determine (1) at both $\Phi _c = 0$ and $\Phi _c = 0.3$, $\beta 1/\beta 5$ contacts were the 
first to rupture, and (2) only for module $C$ was this rupture event {\it invariably} followed by the loss of $\beta 3/\beta 5$ contacts.  When other modules 
ruptured, loss of $\beta 1/\beta 5$ contacts was occasionally followed by loss of the $\beta 1/\beta 2$ strand-pair contacts.

$(\text{Ub})_3$ {\it Must Achieve a Larger} $R_g$ {\it to Rupture at} $\Phi _c = 0.3$:  Figure \ref{fig:21}{\bf A} illustrates the time dependence of $\langle R_g \rangle$ 
at $\Phi _c = 0$ and at $\Phi _c = 0.3$.  Interestingly, the plot reveals that after the second rupture event, the $\langle R_g(\Phi _c,t) \rangle$ increased
more rapidly in the presence of crowders than in their absence.  This is likely a reflection of the fact that at $\Phi_c = 0$ modules $A$ and $C$ were the first
two modules to unfold in 44\% of the trajectories, while at $\Phi _c = 0.3$ these two modules were the first to unfold in only 19\% of trajectories.  Thus, 
$\langle R_g(\Phi _c,t) \rangle$ increased more rapidly in the presence of crowders, because the $R_g$ of $(\text{Ub})_3$ with two adjacent modules unfolded 
is larger than that with two unfolded but non-adjacent modules.  Interestingly, these differences are masked in the time-dependent increase of the end-to-end distance (Fig.\ \ref{fig:21}{\bf B}).
Figure \ref{fig:21}{\bf A} also reveals that the horizontal inflection points marking the third unfolding event occur at different times at $\Phi_c = 0.3$ and in the absence
of crowders.  The difference between these two times was responsible for the difference in average unfolding forces of $\approx 14$ pN illustrated in 
Fig.\ \ref{fig:13}.  Thus, it is clear that despite the highly non-equilibrium nature of this pulling experiment, depletion effects were substantial and it required
a much greater force to reach $f_u$ in the presence of these crowders than in their absence.

We predict that systematic experiments will reveal that polyUb molecules
composed of larger numbers of modules will show greater increases in $\langle f_u(\Phi _c) \rangle$ relative to $\langle f_u(\Phi _c = 0) \rangle$ 
than polyUb molecules composed of fewer repeats.  The size of these differences should increase with decreasing loading
rate.  Finally, it may even be possible to observe differences in the $\langle f_u \rangle$ as a function of unfolding event number (as
in Fig.\ \ref{fig:13}).  The increase in $\langle f_u(\Phi _c) \rangle$, at a fixed $\Phi _c$, for poly Ub is likely to be even more significant for small 
crowding agents as shown in Eq.\ (\ref{eqn:6}) (see below for further discussion).

\section*{Conclusions.}
General theory, based on the concept of depletion effects (see Eqs.\ (\ref{eqn:8}) and (\ref{eqn:6})), shows that
crowding should enhance the stability of proteins,
and hence should result in higher forces to unfold proteins. However,
predicting the precise values of $\langle f_u(\Phi _c) \rangle$ 
is difficult because of the interplay of a number of factors such as
the size of the crowding agents and the number of amino acid residues
in the protein. Despite the complexity a few qualitative conclusions
can be obtained based on the observation that, when only excluded volume interactions
are relevant, then the protein or polyprotein would prefer to be localized in a
region devoid of crowding particles \cite{ThirumalaiPRA88}. The size of such a region $D
\approx  \sigma_c ({\frac{\pi}{6 \Phi_c}})^{\frac{1}{3}}$.
If $D \gg R_g$ then the crowding would have negligible effect on the
unfolding forces. The condition $D \gg R_g$ can be realized by using
large crowding particles at a fixed $\Phi_c$.
In the unfolded state, $R_g \approx 0.2N^{0.6}$ nm \cite{PlaxcoPNAS05} which for Ub leads to
$R_g \approx 2.7$ nm. Thus, $\frac{D}{R_g} = 0.4 \sigma_c$. These
considerations suggest that the crowder with
$\sigma_c = 6.4$ nm would have negligible effect on the unfolding force,
which is in accord with the simulations. On the other hand,
$\frac{D}{R_g} \approx 0.4$ when $\sigma_c = 1$ nm, and hence we expect that
the smaller crowders would have measurable effect on the unfolding
forces. Our simulations are in harmony with this prediction. We expect
that for the smaller crowding agent $\langle f_u(\Phi_c) \rangle$ would scale with
$\Phi_c$ in a manner given by Eq.\ (\ref{eqn:6}).  In general, appreciable effect of crowding on the unfolding forces can be observed only for large
proteins or for polyproteins using relatively small crowding agents.

Although we have only carried out simulations for Ub and (Ub)$_3$ at one non-zero $\Phi _c$, theoretical arguments can be used to predict the changes in $\langle f_u(\Phi _c) \rangle$ as $\Phi _c$ increases.  The expected changes in the force required to unfold a protein can be obtained using a generalization of the arguments of Cheung et al.\cite{Cheung:2005}  In the presence of crowding agents the protein is localized in a region that is largely devoid of the crowding particles \cite{ThirumalaiPRA88}.  The most probable size of the region is $D \sim \sigma_c \Phi _c^{-1/3}$ where $\sigma_c$ is the size of the crowding agent.  If the structures in the DSE are treated as a polymer with no residual structure then the increase in entropy of the DSE upon confinement is $\Delta S / k_B \sim (R_g/D)^{1/\nu}$ where $R_g$ is the dimension of the unfolded state of the protein.  If native state stabilization is solely due to the entropic stabilization mechanism we expect:
\begin{equation}
\label{eqn:6}
\langle f_u(\Phi _c) \rangle \sim T\Delta S/L_c \sim (\frac{R_g}{\sigma_c})^2 \Phi _c^{1/3\nu} (\frac{k_B T}{L_c})
\end{equation}
where $f_u(\Phi _c)$ is the critical force for unfolding the protein, $L_c$ is the gain in contour length at the unfolding transition, and $\nu (\approx 0.588)$ relates $R_g$ to the number of amino acids through the relation $R_g \sim a_D N^\nu$ ($a_D$ varies between 2-4 \AA).  A few comments regarding Eq. (\ref{eqn:6}) are in order.  (1)  The $\Phi _c$ dependence in Eq. (\ref{eqn:6}) does not depend on the nature of the most probable region that is free of crowding particles.  As long as the confining region, which approximately mimics the excluded volume effects of the macromolecule, is characterized by a single length D, we expect Eq. (\ref{eqn:6}) to be valid.  (2)  The additional assumption used in Eq. (\ref{eqn:6}) is that $N \gg 1$, and hence there may be deviations due to finite size effects.  (3)  The equivalence between crowding and confinement breaks down at large $\Phi _c$ values.  Consequently, we do not expect Eq.\ (\ref{eqn:6}) to fit the experimental data at all values of $\Phi _c$.  (4)  It follows from Eq.\ (\ref{eqn:6}) that, for a given $R_g$, small crowding agents are more effective in stabilizing proteins than large ones.  Thus, the prediction based on Eq.\ \ref{eqn:6} is supported by our simulations.  (5)  From the variation of $\langle f_u(\Phi _c) \rangle$ with $\Phi _c$ Ping et al.\ \cite{Ping:2006} suggest that $\langle  f_u \rangle \sim \Phi _c$.  However, the large errors in the measurements cannot rule out the theoretical prediction in Eq. (\ref{eqn:6}).  We have successfully fit their experimental results using Eq. (\ref{eqn:6}) (Fig.\ \ref{fig:21}{\bf C}).  Additional quantitative experiments are required to validate the theoretical prediction.

It is difficult to map the concentrations in g/L used in the study of Ping et al. \cite{Ping:2006} to an effective volume fraction because of uncertainties in the molecular weight of Dextran used in the study.  Hence, a quantitative comparison between theory and experiments is challenging.  A naive estimate may be obtained using the values reported by Weiss et al.\ \cite{Weiss:2004}.  Given that the Dextran used in the study is thought
to have an average molecular weight of 40 kDa and an estimated average hydrodynamic radius of 3.5 nm \cite{Weiss:2004}, we find that $\rho = 300$ g/L corresponds to a volume
fraction $\Phi _c = 0.8$, which is very large.  Nevertheless, $\Phi _c$ must be large when $\rho = 300$ g/L.  
Alternatively, we estimated $\sigma _c/2$ for Dextran using $\sigma _c/2 \approx a_DN^{1/3}$ where $N$ is the number of monomers in a 40 kDa Dextran is 
$40/0.162 \approx 147$.  If the monomer size $a_D \approx 0.4-0.45$ nm, then we find $\Phi_c \approx 0.3-0.4$.
If we assume that $\rho = 300$ g/L corresponds to $\Phi _c \simeq 0.4$ and that
$\langle f_u(\Phi _c) \rangle \sim \Phi_c^{5/9}$ (Eq.\ (\ref{eqn:6})), then at $\Phi _c = 0.3$ we would expect a nearly
18$\%$ increase in $\langle f_u \rangle$.  Similarly, if $\rho = 300$ g/L corresponds to $\Phi _c \simeq 0.3$ then we would expect an
increase in $\langle f_u \rangle$ of approximately 21$\%$.  In any case, we can say that at physiologically relevant volume fractions ($\Phi _c$ $\in $ [0.1,0.3]), 
the percent increase in $\langle f_u \rangle$ is likely to be $\le 20\%$.  Our simulations for $\sigma_c = 1.0$ nm 
predict an increase of 3-4\%, which shows that a more detailed analysis is required to obtain an accurate value of $\sigma_c$ for Dextran 
before a quantitative comparison with experiments can be made.  The larger increase seen in experiments may also be a reflection of the use of 
(Ub)$_8$ rather than a monomer.

Regardless of the crowder size we find that the unfolding pathways are altered in the presence of crowding agents.  It is normally assumed that the rupture of secondary 
structure elements is irreversible if the applied force exceeds a threshold value.  However, when unfolding experiments are carried out in the presence of crowding particles, 
that effectively localize the protein in a smaller region than when $\Phi _c = 0$, reassociation between already ruptured secondary structures is facilitated 
as shown here.  Thus, forced-unfolding cannot be described using one dimensional free energy profiles with $z_u$ as the reaction coordinate \cite{Hyeon:2007}.

We find that the average unfolding force for the final rupture event of the unfolding of $(\text{Ub})_3$
occurred at much larger values in the presence of crowders than in their absence.  With $\sigma _c = 6.4$ nm, which has practically no effect on the unfolding force of the 
monomer, and $\Phi _c = 0.3$ even with unfolding of two modules 
the interactions between the stretched modules and protein are small ($D \approx 1.2\sigma_c$).  Only upon unfolding of the third Ub do crowding effects become 
relevant, which leads to an increase in $\langle f_u(\Phi _c) \rangle$.  Our results suggest that
$\langle f_u(\Phi _c) \rangle / \langle f_u(0) \rangle$ should increase with the number of modules in the array and
that it may be possible to detect differences in $\langle f_u \rangle$ which are conditional on the unfolding event number.  We speculate that naturally 
occurring polyproteins that are subject to mechanical stress have evolved to take advantage of precisely such enhanced depletion effects.

\section*{Acknowledgments}
This work was supported by a grant from the National Science Foundation (CHE 05-14056).  DLP is grateful for a Ruth L. Kirschstein Postdoctoral Fellowship from the National Institute of General Medical Sciences (F32GM077940).  Computational time and resources for this work were kindly provided by the National Energy Research Scientific Computing (NERSC) Center.

\providecommand{\refin}[1]{\\ \textbf{Referenced in:} #1}

\newpage
\section*{\bf Tables}

\begin{table}[htdp]
\caption{Module Unfolding Order Frequencies at $\Phi _c = 0$ and $\Phi _c = 0.3$.}
\begin{center}
\begin{tabular}{|c|c|c|}
\hline 
Unfolding Order & Frequency Observed at $\Phi _c = 0$ & Frequency Observed at $\Phi _c = 0.3$\\
\hline \hline
$C \rightarrow B \rightarrow A$ & 0.44 &  0.56\\
\hline
$C \rightarrow A \rightarrow B$ & 0.38 &  0.13\\
\hline
$B \rightarrow A \rightarrow C$ & 0.00 &  0.00\\
\hline
$B \rightarrow C \rightarrow A$ & 0.13 &  0.25\\
\hline
$A \rightarrow B \rightarrow C$ & 0.00 &  0.00\\
\hline
$A \rightarrow C \rightarrow B$ & 0.06 &  0.06\\
\hline
\end{tabular}
\end{center}
\label{tbl:moduleOrderFrequencies}
\end{table}

\newpage
\section*{Figure Captions.}

\begin{figure}[ht]
\includegraphics[width=7.00in]{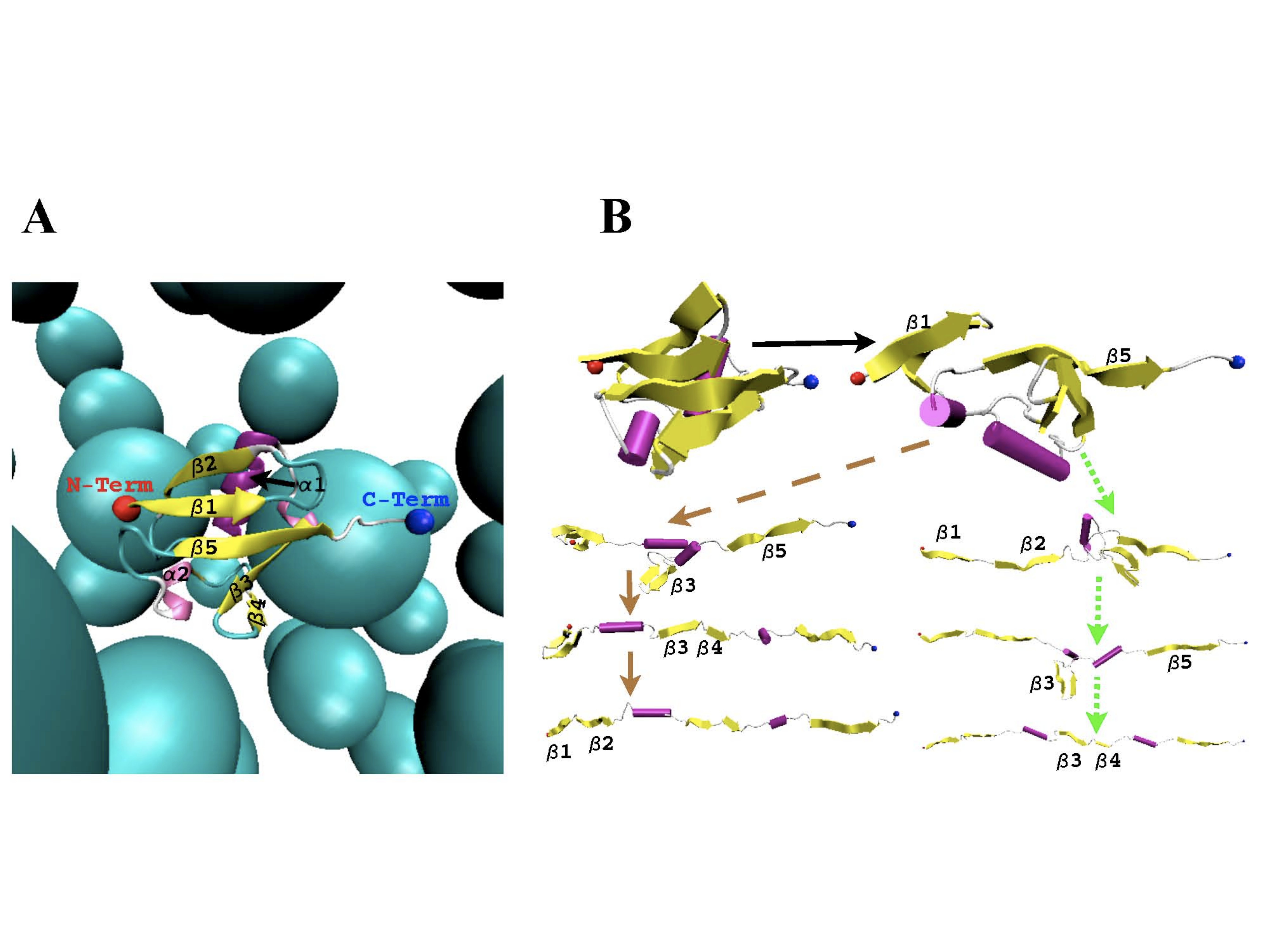}
\caption{}
\label{fig:1}
\end{figure}

Fig.~\ref{fig:1}: { }({\bf A}) Cartoon representation of the native structure of ubiquitin (PDB accession id 1UBQ) in the presence of spherical crowding agents.  The five beta-strands, 
labeled $\beta $1 through $\beta $5, are colored in yellow.  The two alpha-helices ($\alpha $1 and $\alpha $2) are shown in 
purple.  The N- and C-terminal beads are represented as spheres.  In our simulations the N-terminal bead was held fixed while the C-terminal bead
was pulled via a tethered spring.  ({\bf B}) Snapshots from an unfolding trajectory illustrating the main ubiquitin (Ub) unfolding pathway (brown-dashed arrows) and an alternate
unfolding pathway (green-dotted arrow).  In both pathways the initial unfolding event corresponds to separation of the C-terminal strand $\beta $5 from the N-terminal strand $\beta $1.  Along the main pathway, this is quickly followed by separation of $\beta $5 from $\beta $3.  The penultimate rupture event along the main pathway corresponds to disruption of the $\beta $3/$\beta $4 strand-pair, while the N-terminal $\beta $1/$\beta $2 strand pair is the last to break.  The trajectory illustrated here was generated
at $\Phi _c = 0.0$ and $r_f = 160 \times 10^3$ pN/s. An alternate pathway was observed at $\Phi _c = 0.0$ and $r_f = 4 \times 10^3$ pN/s.  Along this pathway separation of $\beta $5 from $\beta $1 is followed by separation of $\beta $1/$\beta $2.   The final two rupture events correspond to those of the $\beta $5/$\beta $3 contacts and $\beta $3/$\beta $4 strand-pair respectively.  (Figures generated  with VMD \cite{Humphrey:1996})

\begin{figure}[ht]
\includegraphics[width=7.00in]{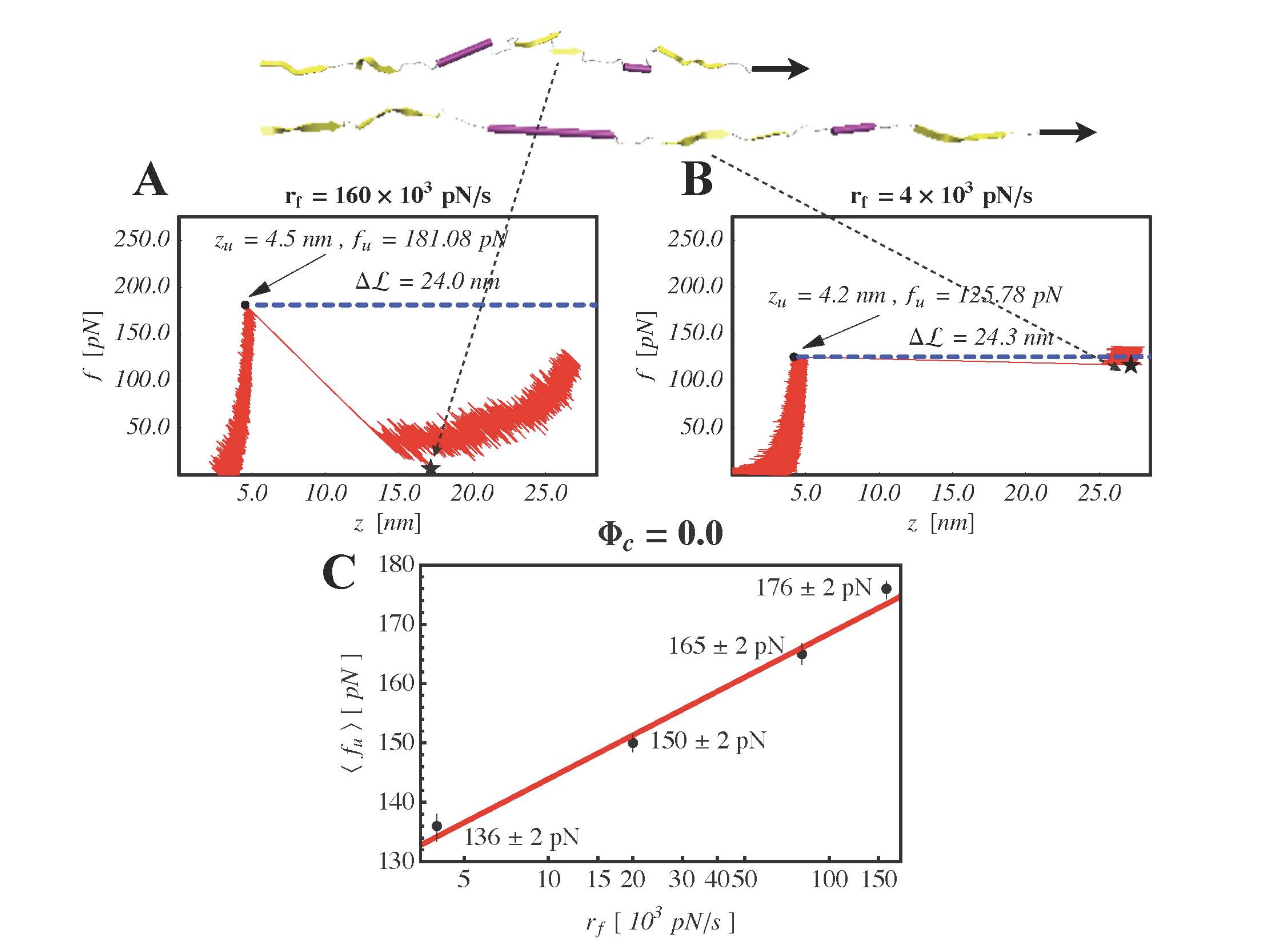}
\caption{}
\label{fig:2}
\end{figure}

\noindent Fig.~\ref{fig:2}: { }Force-extension curves (FEC's) at two different loading rates, ({\bf A}) $r_f =160 \times 10^3$ pN/s, 
and ({\bf B}) $r_f = 4 \times 10^3$ pN/s.  Data from the simulation is presented as a red trace.  For each trajectory a black arrow points
to the unfolding force, $f_u$.  $z_u$ corresponds to the extension of the molecule along the pulling (i.e., z-) direction evaluated at $f_u$.  
$\Delta \mathcal{L}$ (dotted blue line) is the contour length increment, and is a measure of the amount of chain released
in an unfolding event.  We measured $\Delta \mathcal{L}$ as $\mathcal{L} - z_u$, where $\mathcal{L} = (N-1) \times \sigma = 75 \times 0.38$ nm 
is a nominal contour length of 28.5 nm.  Stars in each subfigure mark the minimum force observed after an unfolding event, and chain
conformations corresponding to the starred points in figures {\bf A} and {\bf B} are illustrated at the center of the figure.  
Unfolding events at smaller $r_f$ resulted in larger molecular extensions before
significant resistance was encountered.  (Yellow arrows correspond to beta-strands and purple cylinders correspond to $\alpha $-helices).  (Figures generated with VMD \cite{Humphrey:1996}).  ({\bf C}) $\langle f_u \rangle$ vs.\ $r_f$ evaluated at $\Phi _c = 0$.  The red curve corresponds to a
linear-least squares fit to the set of basis functions $\{1, \ln(r_f)\}$ and demonstrates that $\langle f_u \rangle \sim \ln(r_f)$. (Note that the abscissa is a log-scale).  
Each point is labeled Mean $\pm $ standard error.  Statistics at $r_f = 160 \times 10^3$ pN/s, $80 \times 10^3$ pN/s, $20 \times 10^3$ pN/s,
and $4 \times 10^3$ pN/s were calculated from 50, 49, 50, and 16 trajectories respectively. 

\begin{figure}[ht]
\includegraphics[width=8.00in]{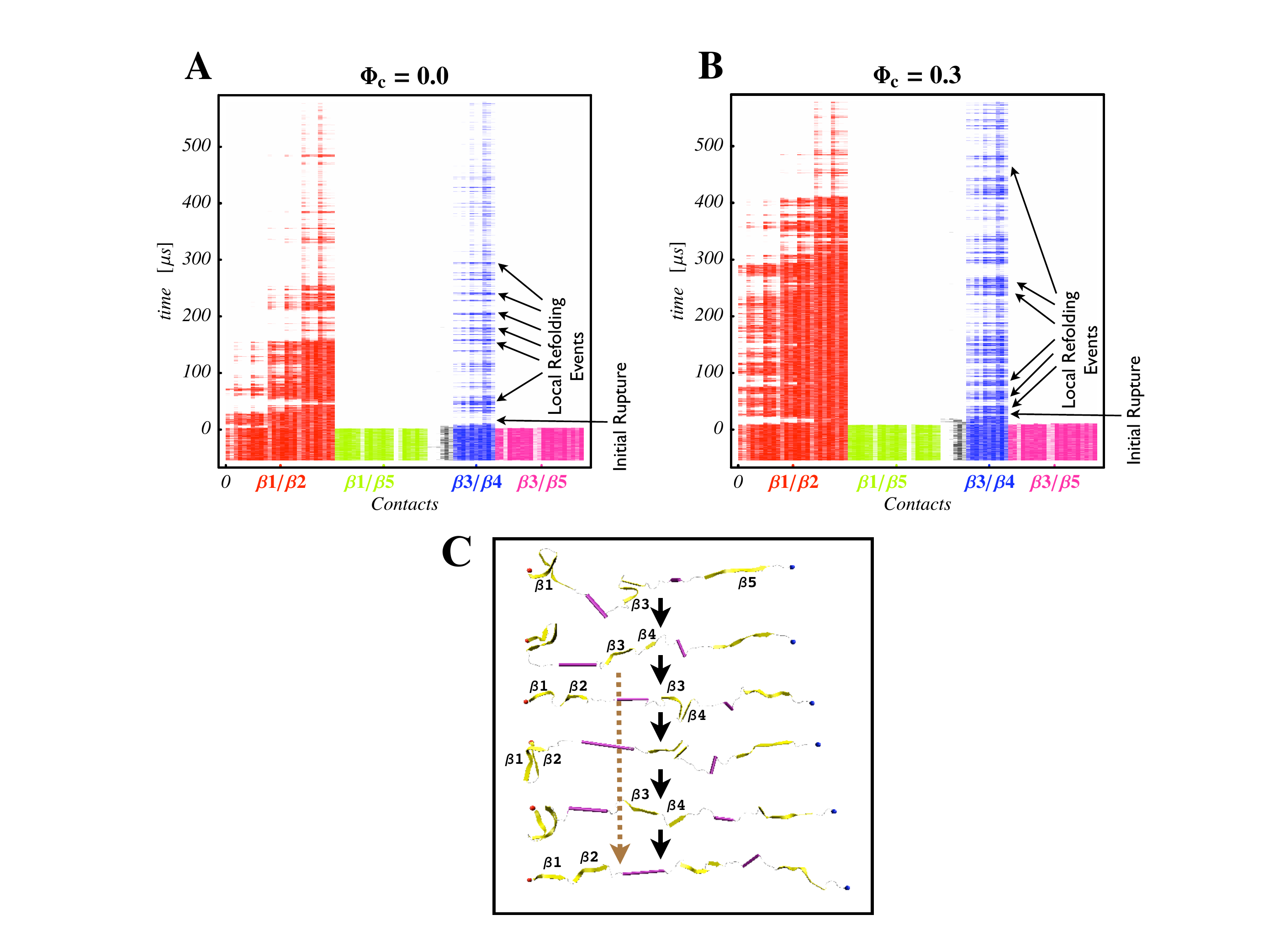}
\caption{}
\label{fig:12}
\end{figure}

\noindent Fig.~\ref{fig:12}: { }Rupture events at $\Phi _c = 0.0$ ({\bf A}) and at $\Phi _c = 0.3$ and $\sigma _c = 6.4$ nm ({\bf B}).  The figure illustrates that after the initial rupture event at step 0, subsequent unfolding was affected by the crowding agent.  $\beta $1/$\beta $2 contacts and $\beta $3/$\beta $4 persisted longer and there were many more local refolding
events in the presence of the crowders ({\bf B}) than in their absence ({\bf A}). ({\bf C})  Snapshots from an unfolding trajectory at $\Phi _c = 0.3$ and $r_f = 4 \times 10^3$ 
pN/s illustrate the primary difference between unfolding at $\Phi _c = 0.3$ and at
$\Phi _c = 0.0$.  The brown-dotted arrow shows unfolding without any local-refolding events, while the sequence of black arrows illustrates local-rupture and -refolding events.  The hallmark of forced-unfolding in a crowded environment is repeated breaking and reforming of contacts after an initial rupture event.  (Figures generated  with VMD \cite{Humphrey:1996})

\begin{figure}[ht]
\includegraphics[width=8.00in]{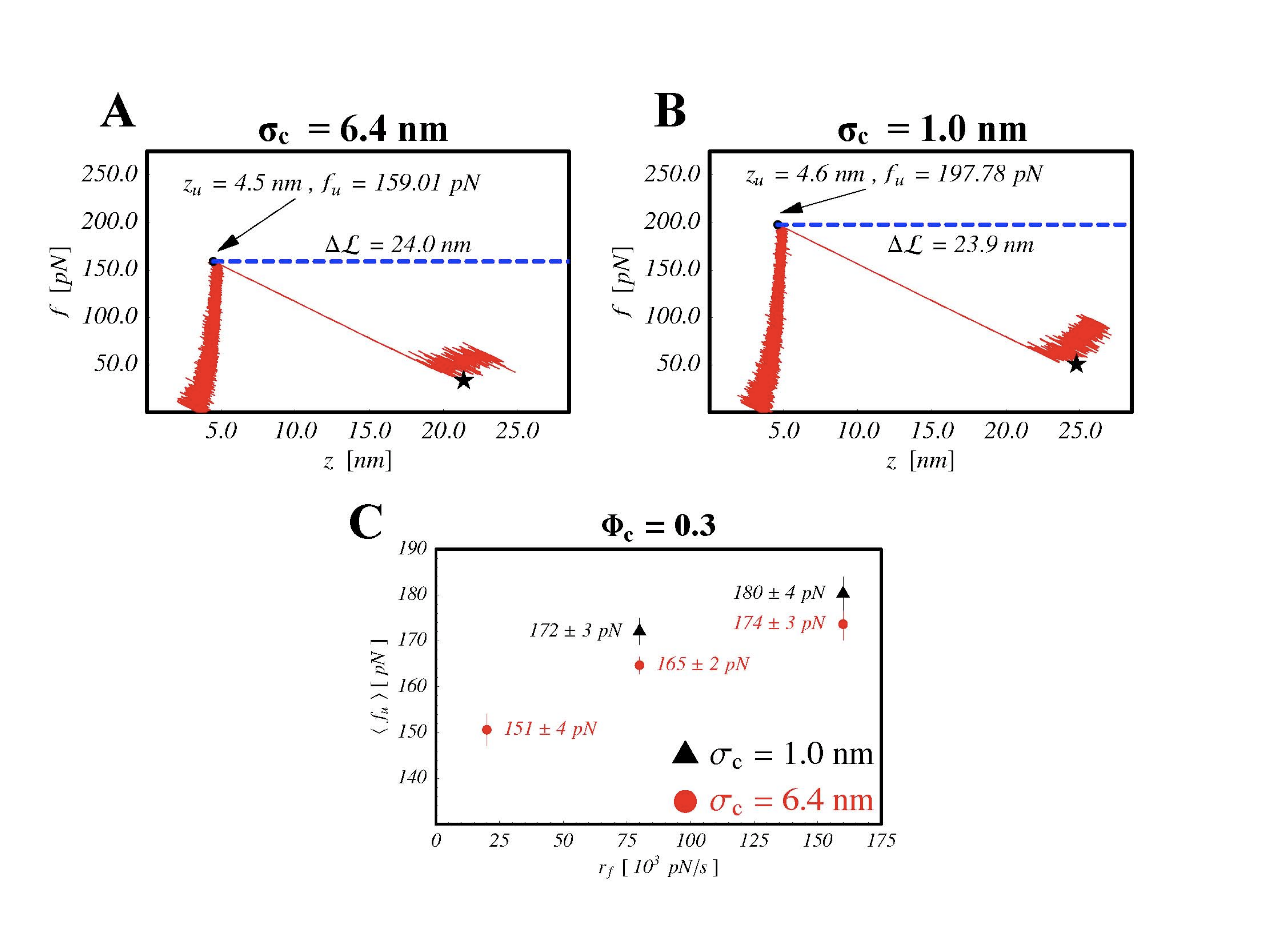}
\caption{}
\label{fig:9}
\end{figure}

\noindent Fig.~\ref{fig:9}: { }({\bf A}) Examples of force-extension traces resulting from simulation in the presence of spherical crowding agents of diameter
$\sigma _c = 6.4$ nm and obtained at $r_f = 80 \times 10^3$ pN/s. ({\bf B}) Examples of force-extension traces resulting from simulation 
in the presence of spherical crowding agents of diameter $\sigma _c = 1.0$ nm and obtained at $r_f = 80 \times 10^3$ pN/s.  
Both subfigures are labeled as in Fig.\ \ref{fig:2}({\bf A} and {\bf B}).  ({\bf C}) $\langle f_u \rangle$ vs.\ $r_f$ evaluated at $\Phi _c = 0.3$.  Black triangles and red circles
correspond to spherical crowders of diameter $\sigma _c = 1.0$ nm and $\sigma _c = 6.4$ nm respectively.  Each point is labeled Mean $\pm $
standard error.  Statistics for each point were calculated from 16 independent trajectories.  Only the $\sigma _c = 1.0$ nm had an appreciable effect
on $\langle f_u \rangle$ when compared to those obtained at identical $r_f$ and at $\Phi _c = 0.0$ (see Fig.\ \ref{fig:2}{\bf C})

\begin{figure}[ht]
\includegraphics[width=7.00in]{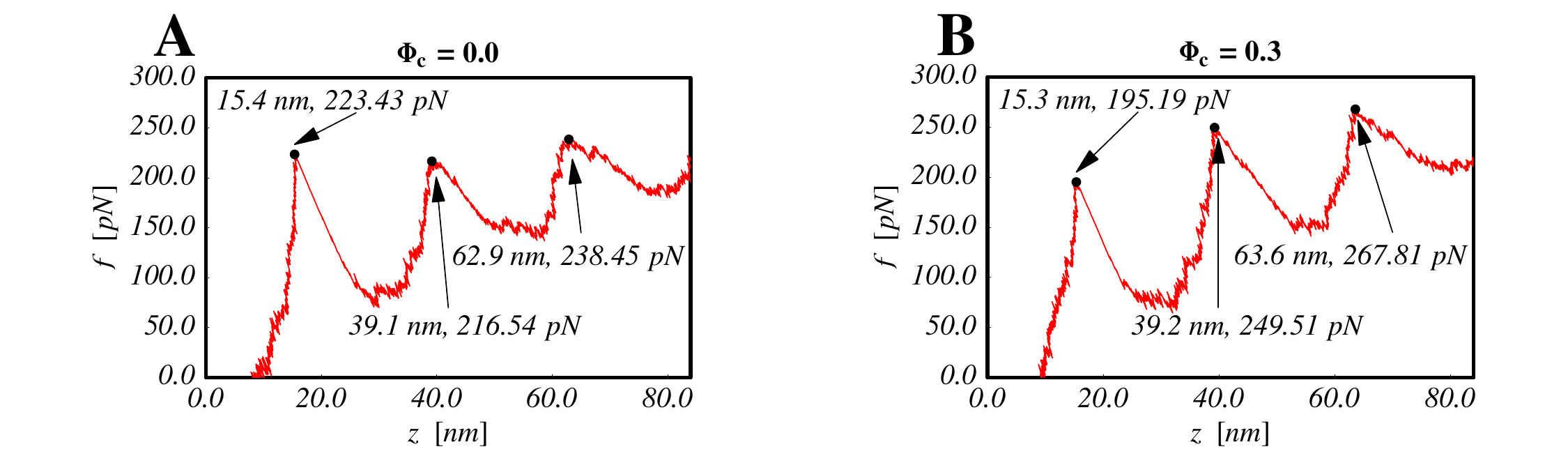}
\caption{}
\label{fig:14}
\end{figure}

\noindent Fig.~\ref{fig:14}: { }FEC's for $(\text{Ub})_3$ forced unfolding at $\Phi _c = 0.0$ ({\bf A}) and at $\Phi _c = 0.3$ ({\bf B}).  Trajectories were
generated at $r_f = 640 \times 10^4$ pN/s and with crowders of diameter $\sigma _c = 6.4$ nm.  Black arrows mark each trajectory's three unfolding events.

\begin{figure}[ht]
\includegraphics[width=7.00in]{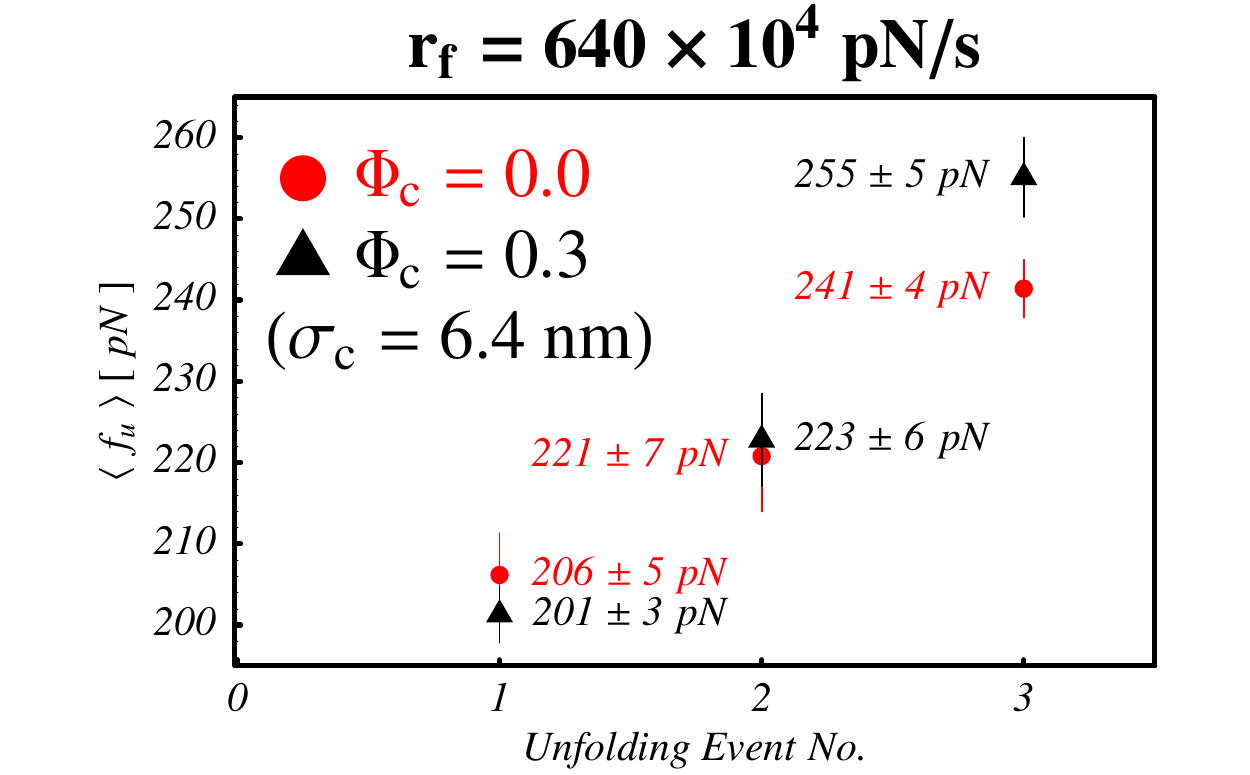}
\caption{}
\label{fig:13}
\end{figure}

\noindent Fig.~\ref{fig:13}: { }$\langle f_u \rangle$ vs.\ unfolding event number for the unfolding of $(\text{Ub})_3$ 
in the presence of spherical crowders of diameter $\sigma _c = 6.4$ nm ($\Phi _c = 0.3$, black triangles) and in their absence ($\Phi _c = 0.0$, red circles).  
The individual modules of the polyUb tandem were N-C-linked and the loading rate was $640 \times 10^4$ pN/s.  Each point
is labeled Mean $\pm$ standard error.  Statistics for each point were calculated from 16 independent unfolding trajectories.  An unfolding event corresponded
to the unfolding of an individual module.  Note that although the crowders had little effect on $\langle f_u \rangle$ 
for the first and second unfolding events, they had a substantial effect on the last unfolding event.  $\langle f_u \rangle$ increased by $\approx 14$ pN for the 
last unfolding event.

\begin{figure}[ht]
\includegraphics[width=7.00in]{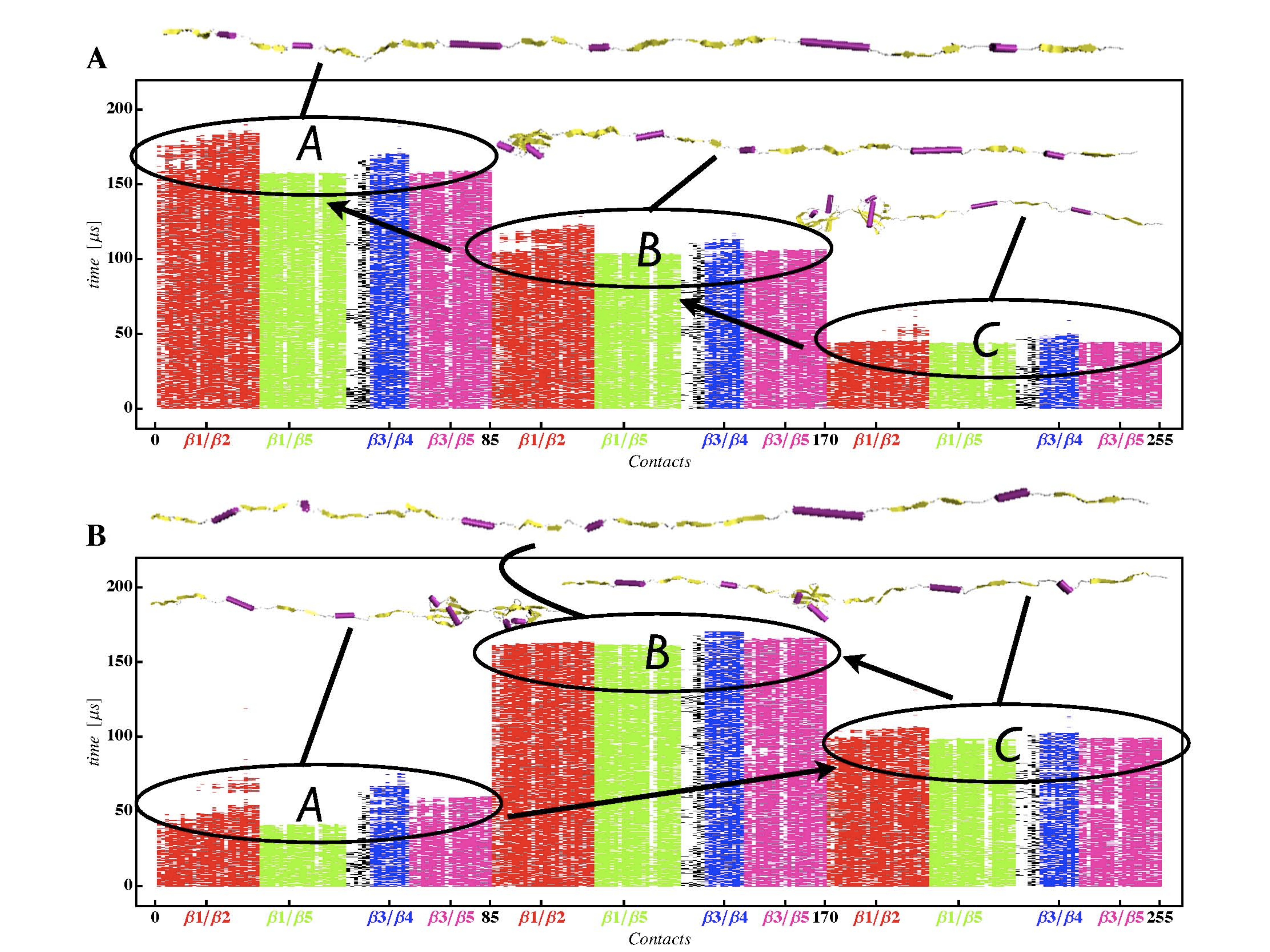}
\caption{}
\label{fig:20}
\end{figure}

\noindent Fig.~\ref{fig:20}: { }Illustration of the stochastic nature of module unfolding.  In ({\bf A}) module $C$ (the most proximal to the applied force) unfolds first, followed by module $B$, and finally by rupture of module $A$.  In ({\bf B}) the order of module unfolding events is $A \rightarrow C \rightarrow B$.  The two trajectories illustrated here were both collected at 
$\Phi _c = 0$ and $r_f = 640 \times 10^4$ pN/s.  Table \ref{tbl:moduleOrderFrequencies} provides the frequencies at which the different possible orders were observed at $\Phi _c = 0$ and at $\Phi _c = 0.3$.  The diameter of the crowders is $\sigma_c = 6.4$ nm.

\begin{figure}[ht]
\includegraphics[width=7.00in]{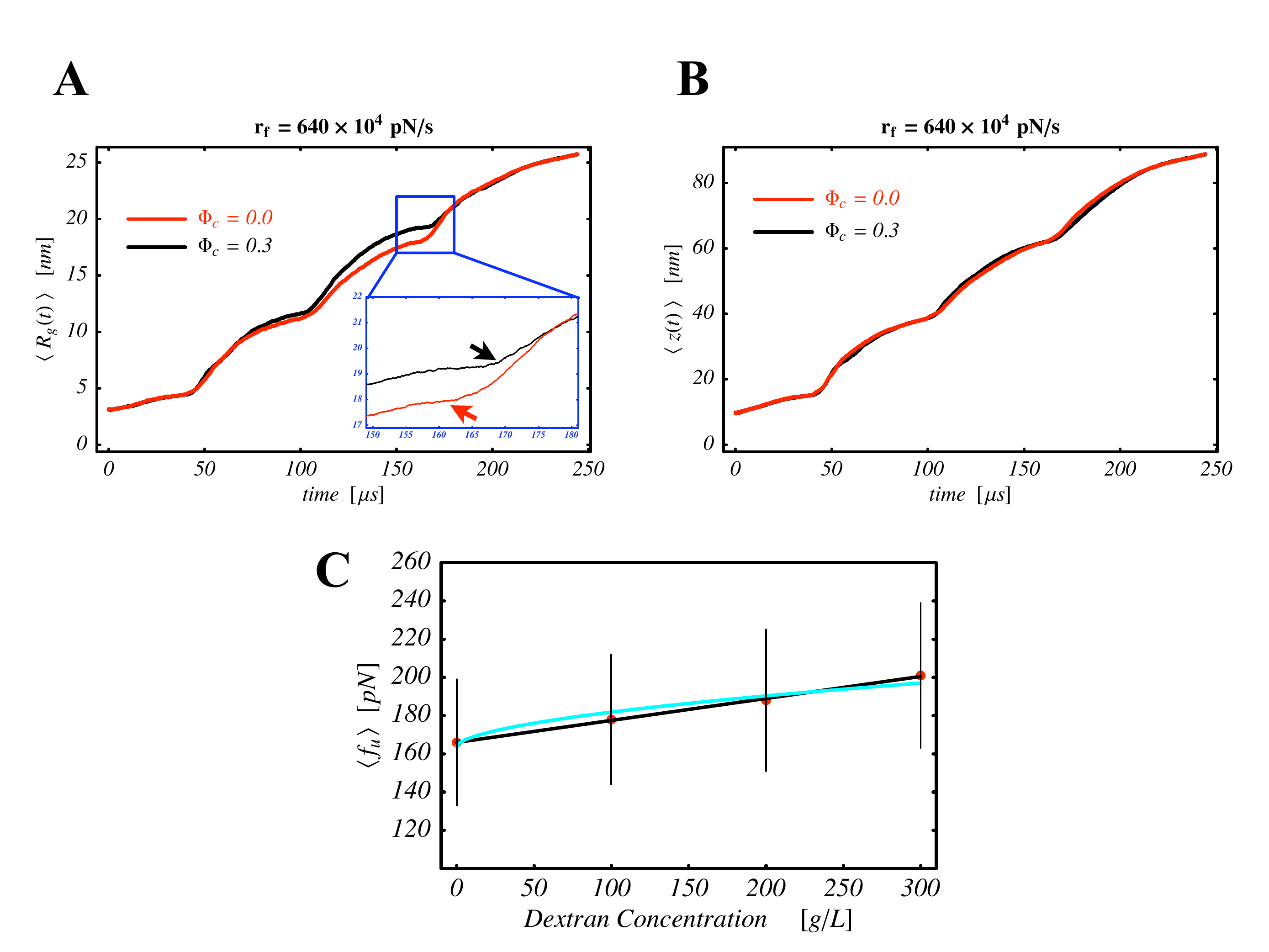}
\caption{}
\label{fig:21}
\end{figure}

\noindent Fig.~\ref{fig:21}: { }({\bf A}) $\langle R_g(t) \rangle$ versus time at $\Phi _c = 0$ (red) and at $\Phi _c = 0.3$ (black).  Although the $\langle R_g \rangle$ increases more rapidly with time after the second rupture event at $\Phi _c = 0.3$ than at $\Phi _c = 0$, the inset reveals that a larger $R_g$ must be achieved to initiate the final rupture event in the presence of crowding particles with $\sigma_c = 6.4$ nm.  (See text for additional discussion).  ({\bf B}) $\langle z(t) \rangle$ versus time at $\Phi _c = 0$ (red) and at $\Phi _c = 0.3$ (black).  
$z(t)$ cannot discriminate between unfolding at the different volume fractions and hence is less suitable (than $R_g$) as a potential reaction coordinate.  ({\bf C}) The experimental results of Ping et al.\ \cite{Ping:2006} (red circles) for unfolding-force, 
$\langle f_u \rangle$ as a function of Dextran concentration, $\rho$.  The black line is a fit assuming $\langle f_u \rangle \sim \rho$ and the blue assuming $\langle f_u \rangle \sim \rho^{5/9}$.  Although both fits are consistent with the data, based on theoretical considerations (see Eq.\ (\ref{eqn:6})) we prefer the blue fit (see text).  (Standard deviations taken from Ping et al.\ \cite{Ping:2006})
\newpage

\end{document}